HCist – International Conference on Health and Social Care Information Systems and Technologies 2024

# WhatsApp as an improvisation of health information systems in Southern African public hospitals: A socio-technical perspective

Meke Kapepo[a]* , Jean-Paul Van Belle[b], Edda Weimann[c]

*Department of Information Systems,University of Cape Town, 7701, South Africa*

**Abstract**

Digital health interventions, particularly electronic referrals (e-referrals) and health information systems, have revolutionised clinical workflows in public hospitals by automating processes. However, the utilization of e-referrals has yielded mixed outcomes, with varying levels of success in organisational processes. This paper explores improvisation of health information systems in Southern African public hospitals from a socio-technical perspective. In particular, the paper explains the design-reality gaps giving rise to improvisations of mandated health information systems in order to understand their occurrence and impact on referral outcomes. We employed the design-reality framework and the Process framework for Healthcare Information System Workarounds and Impacts to explain the socio-technical issues related to the phenomenon of interest. We conducted semi-interviews with 31 respondents from health organisations as case studies. Respondents from two public hospitals in South Africa and two in Namibia were interviewed to examine how they devised improvisations to various health information systems in each setting. The findings indicated that use of WhatsApp in the absence of mandated health information systems (HIS) or improvisation of existing HIS was reported to be advantageous for healthcare practitioners (HCPs), leading to improved efficiency and productivity in the execution of referral activities. Additionally, HCPs reported positive outcomes for health organisations related to continual professional development in the given settings. The findings further show a relationship between design-reality gaps and improvisations enacted by HCPs in both case studies. The observed gaps are related to poor management systems and structures, lack of HCPs' involvement in the roll-out of HIS and inadequacies of existing HIS to support referral tasks. These study findings can be insightful and useful to system developers and other stakeholders for devising measures to address the gaps. Improvisations and use of WhatsApp for healthcare service delivery present opportunities for innovation and process improvements for health organisations.

* Corresponding author. E-mail address:meke.kapepo@uct.ac.za

## 1. Introduction

Information systems can be leveraged for intended and unintended purposes, leading to positive or negative consequences within healthcare organisations.

While the intended use refers to following the designed purpose and objectives of the system [1], unintended use involves tasks arising from constraints or misfits with that design [1, 2]. Research suggests that unintended use, often characterized by workarounds or Shadow Information Technology (like using WhatsApp for communication), can emerge despite the potential benefits of information systems [3, 4,5]. This study examines improvisations to health information systems (HIS) used for facilitating patient referrals. In the information systems discipline, this phenomenon is widely observed, and is characterized as workarounds, feral information technology (Feral IT), Shadow Information Technology (Shadow IT), bricolage or improvisation [6, 7]. In this study, we use the term improvisations to define workarounds to mandated HIS and the use of WhatsApp as shadow information technology. These improvisations, while addressing the system inadequacies, can introduce security risks due to their lack of official sanction [6, 7, 8]. Despite these drawbacks, workarounds can also expose areas for improvement in the original system design [6, 9, 10]. However, this phenomenon remains under-theorized in healthcare, with limited research exploring the link between unintended use of health information systems and the socio-technical factors influencing such workarounds, particularly in resource-constrained settings [11, 12].To fill this gap, we seek to examine the design-reality gaps giving rise to workaround practices to mandated health information systems in order to understand their occurrence and impact on referral outcomes. We conducted a multi-case study at four tertiary public hospitals: two situated in the Western Cape Province in South Africa and two in the Khomas region in Namibia. To understand these improvisations, we interviewed 31 healthcare providers to gain their perspectives in the given settings. Thematic analysis of the interviews revealed that healthcare practitioners experienced inadequacies with the mandated health information systems (HIS), as a result, they improvised their referral activities using WhatsApp and other paper-based methods. These inadequacies revealed gaps between the design of the existing HIS and realities of healthcare practitioners. These gaps led healthcare providers to adopt WhatsApp as a workaround, finding it more efficient and improving productivity in the execution of referral activities. Interestingly, this improvisation also fostered improved communication and efficient collaboration among healthcare practitioners.

Our findings further show a relationship between design-reality gaps and improvisations enacted by HCPs in both case studies. We identified other contributing factors, including poor management systems and structures, lack of healthcare provider involvement in HIS implementation, and inadequacies in supporting referral workflows. These insights can inform stakeholders, particularly system developers, to bridge the design-reality gaps. By addressing these issues, future HIS implementations can optimize referral processes and potentially improve patient care delivery. This paper is structured as follows: Section 2 provides a background to contextualize the research. Section 3 outlines the theoretical frameworks employed. The methodology adopted is then described in Section 4, followed by a discussion of the research findings in Section 5. Finally, Section 6 highlights the implications of the research.

## 2. Background

Electronic patient referrals (e-referrals) are commonly studied in high-income country contexts. More recently, e-referrals have been popularly used during the COVID-19 pandemic to ensure effective and efficient health service delivery in different settings [14, 15]. The extensive and successful use of e-referrals is reported to be linked to effective and efficient referral processes in health organisations. Electronic referrals are usually designed as a module that comes as part of HIS software solution. These HIS have been used in intended and unintended ways which can result in positive or negative outcomes in the organisational process. The intended use of an information system refers to its use for the designed and planned objectives [1]. On the other hand, unintended use involves tasks or activities that were not part of the original purpose and design of the system [4, 5]. The benefits derived from the use of these systems were the improvement of communication and collaboration between HCPs and specialists [13], a streamlined referral process and a seamless exchange of information [15, 16, 17], a decrease in unnecessary visits to specialists and decreased waiting times [18, 19]. Literature also reports that electronic referral solutions are used in these countries to provide specialized services to health institutions in remote areas. Most of the services recorded are both synchronous and asynchronous for teleconsultations and are also used to seek a second opinion on health cases. Despite the documented benefits of e-referrals in the global north, limited literature on the use of electronic referrals exists in low- and middle-income countries (LMIC).

A few scholars have documented the successful use of e-referrals in medically underserved settings. LMICs experience constraints related to implementation issues, infrastructural challenges and unintended use of e-health applications in health organisations [2, 12, 13].

This study employs a multiple case study strategy to examine the phenomenon of interest in public hospitals situated in Southern Africa. Two healthcare organisations, namely the Western Cape Department of Health in South Africa and the Ministry of Health and Social Services in Namibia, were selected given their shared similarities in their referrals systems. Additionally, e-referral interventions were implemented in both organisations with an aim of streamlining the referral process. Both healthcare systems render healthcare services through the public and private sectors [20, 21]. In both settings, the majority of the population (90%) is served by the public health system administered by the national government and provincial or regional health departments [21, 22, 23]. Healthcare services are delivered through various health facilities, including district, regional, tertiary, central, and specialized hospitals, as well as primary healthcare centres and clinics.

The Western Cape Province serves a population of approximately 6 million, where healthcare services are provided through three tertiary hospitals, forty-two districts, five regional hospitals, and twenty-two clinics [21]. Conversely, Namibia, with a population of approximately 2.5 million, has 706 health facilities rending healthcare services to 14 regions [20, 22]. Despite the relatively high number of hospitals in Namibia, access to healthcare remains a challenge in remote regions, because services are concentrated in urban areas, leaving underserved areas with limited access to healthcare services.

Both healthcare systems encounter challenges related to self-referrals, incomplete and inefficient referral services [13, 20], prompting the implementation of an m-health intervention called Vula [2]. This intervention was quickly adopted by healthcare practitioners; however, it was also reported that there was a delay in the responses from practitioners on referral messages sent through the mobile application [35]. The significance of the multiple cases lies in the need for additional inquiry into the underlying factors contributing to response delays in the referral messages on the Vula platform and the partial adoption of this intervention in Namibia. The theoretical perspective of the study is discussed in the next section.

## 3. Theoretical perspective of the study

We employed the design-reality gap model to explain the rationales behind improvisations and their relationship to design-reality dimensions. The Design-reality gap framework describes the difference between the "design" or assumptions built into the IT artefact by designers and the reality of the real needs of end-users in the given context [24]. The design-reality gap (DRG) framework is a robust model for explaining multiple cases of HIS success and failure [24, 25, 26] . The design-reality framework was coined by Heeks [24] as an analytical framework for evaluating and predicting the success and failure of health information systems (HIS). In his study, he sought to understand the success and failure of health information systems with a focus on the contingent realities of healthcare institutions in developing contexts. The original model consists of seven dimensions namely, information, technology, processes, objectives and values, staffing and skills, management systems and structures and other resources. This model was later extended with additional constructs as shown in figure 1.

| Actuality | | Design | Constructs |
|---|---|---|---|
| Information | ↔ | Information | Original constructs |
| Technology | ↔ | Technology | |
| Processes | ↔ | Processes | |
| Objectives and values | ↔ | Objectives and values | |
| Staffing and skills | ↔ | Staffing and skills | |
| Management systems and structures | ↔ | Management systems and structures | |
| Other resources | ↔ | Other resources | |
| Privacy and security | ↔ | Privacy and security | Extended constructs |
| Ethical issues | ↔ | Ethical issues | |

Fig 1: Extended design-reality gap model : Source [27]

The DRG framework has been used in many settings; and other scholars have expanded this model with additional constructs. For example, Masiero [25] included the concept of "other factors" where she explored causal factors for success and failure of information systems. Makoza and Albertus [26] , explored a digital contact tracing mobile application (App) a COVID Alert SA App (CASAA). In their study, they expanded the DRG model with additional constructs on privacy, security, and ethical issues. This study draws on the above constructs of the design-reality gap framework to understand the causal factors of HCPs for enacting workaround practices. Furthermore, the notion of contingency is relevant in this study, as fit or congruence looks at the mismatch and match between systems design and systems delivery. The aim is to look at alternative ways to adapt the health information systems to avoid mismatches. Furthermore, the dimensions are applied to illustrate the design-reality gaps between technology, people, and organisations [26, 27] . In this study, Makoza and Albertus's [27] version of the model is adopted, as it includes additional constructs on privacy; security and ethical issues which are particularly relevant to this study which includes patient health data. The research methodology we adopted is discussed in the next section.

## 4. Methodology

We adopted an interpretive perspective to investigate how healthcare providers (HCPs) enacted improvisations to the mandated health information systems (HIS) in South African and Namibian hospitals. The interpretive paradigm is premised on reality that is socially constructed and it is suited to examining socio-technical practices of health care providers in the given settings [27]. This paradigm is therefore appropriate for examining HCPs practices related HIS improvisation in the given settings. Employing a multi-case study strategy, we selected hospitals in both countries to explore this phenomenon within its real-life context [29]. Specifically, we purposively sampled the Western Cape Department of Health in South Africa and the Ministry of Health and Social Services in Namibia, focusing on the implementation of health information systems in both settings. Within these cases, we examined the embedded cases—the Western Cape Referral System (hospitals A and B) and the Khomas Referral Systems (hospitals C and D)—to gain insights into the socio-technical practices of healthcare providers. Semi-structured interviews were conducted with 31 participants from the selected hospitals, complemented by observations of referral activities on the HIS platform, to deepen our understanding of these practices [30].

Table 1. Demographic profile of participants

| | Case Study 1 | | Case Study 2 | |
|---|---|---|---|---|
| **Jon title** | **Participant** | **Code** | **Participant** | **Code** |
| **Head of unit** | Participant 1 | PT_01 | Participant 14 | PT_14 |
| | Participant 10 | PT_10 | Participant 18 | PT_18 |
| **Medical officers** | Participant 2 | PT_02 | Participant 15 | PT_15 |
| | Participant 3 | PT_03 | Participant 16 | PT_16 |
| | Participant 4 | PT_04 | Participant 17 | PT_17 |
| | Participant 5 | PT_05 | Participant 19 | PT_19 |
| | Participant 6 | PT_06 | Participant 20 | PT_20 |
| | Participant 7 | PT_07 | Participant 21 | PT_21 |
| | Participant 8 | PT_08 | Participant 22 | PT_22 |
| | Participant 9 | PT_09 | Participant 23 | PT_23 |
| | Participant 11 | PT_10 | Participant 24 | PT_24 |
| | Participant 12 | PT_12 | Participant 26 | PT_26 |
| | Participant 13 | PT_13 | Participant 28 | PT_28 |
| **Nurses** | | | Participant 27 | PT_27 |
| | | | | PT_29 |
| **IT Personnel** | Participant 31 | PT_31 | Participant 25 | PT_25 |
| | | | Participant 30 | PT_30 |

The study results are reported using the code refences to indicate and demonstrate instances of the data in the interview transcripts. The study findings are presented in a form of themes consisting of code references from text files, which serve to illustrate instances of data, and the relationships between concepts from the chosen theoretical frameworks employed in the research [31, 32]. We present and demonstrate the relationship between the themes discovered in the empirical data in the matrix table in appendix A. The study findings are discussed in the next section.

## 5. Findings

Findings from both cases show various improvisations were enacted due to design-reality gaps experienced with HIS implemented in each case. In both cases, an electronic application was implemented with the objective of facilitating and transmitting patient referral information. However, in both cases, the applications designed and deployed in both cases were a mismatch with referral tasks. As a result, HCPs were enacting improvisations in the form of fitting, augmenting and workarounds as shown in figure 2.

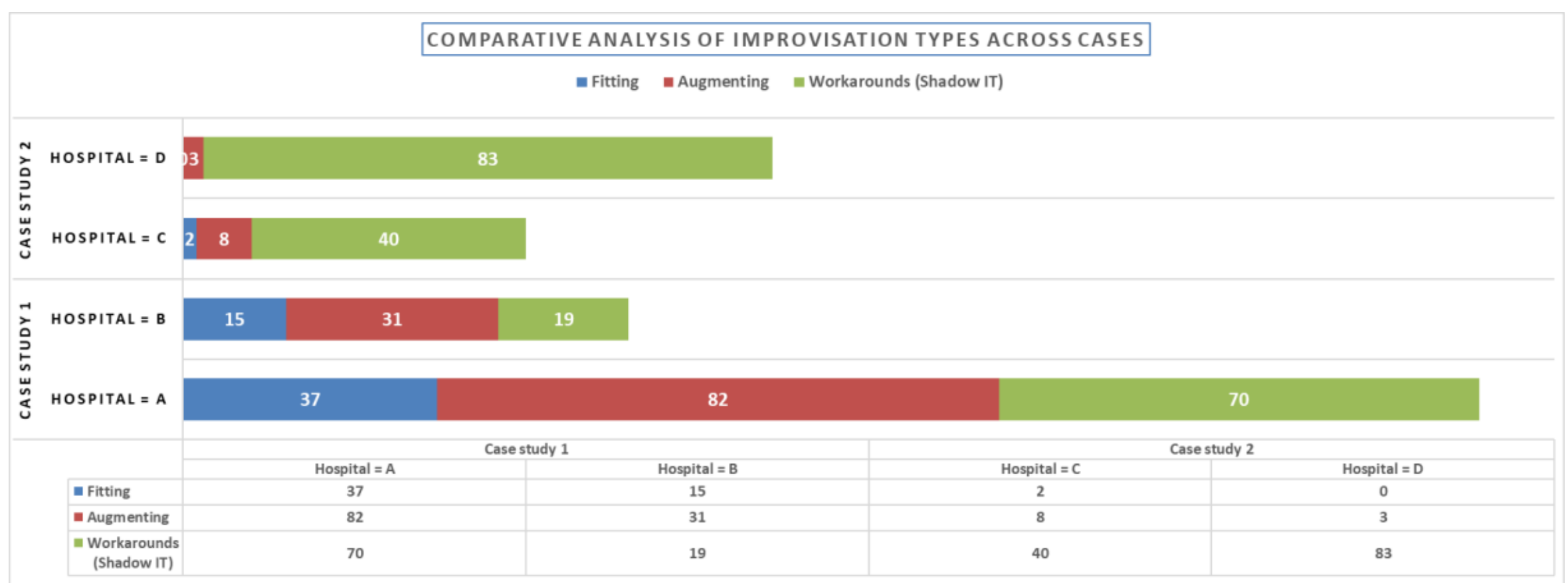

| | Case study 1 | | Case study 2 | |
|---|---|---|---|---|
| | Hospital = A | Hospital = B | Hospital = C | Hospital = D |
| Fitting | 37 | 15 | 2 | 0 |
| Augmenting | 82 | 31 | 8 | 3 |
| Workarounds (Shadow IT) | 70 | 19 | 40 | 83 |

Fig. 2. Case study 1: Code references to Design-reality dimensions across case

Fitting involves undertaking a task of changing the structure of work or computing to accommodate technical misfits, while augmenting undertaking additional work to make up for a misfit between an IT artefact and user needs. Workarounds are forms or improvising or adapting processes or IT artefacts to accommodate misfits. In this study, various improvisations were identified and coded to the design-reality gap model. We describe the positive impacts of these improvisations in the next section.

### *5.1. Positive impacts of Improvisations*

Empirical evidence shows that augmenting mandated systems and processes or using WhatsApp as the primary tool for communication in the absence of an official system has proven to be advantageous for healthcare practitioners (HCPs), leading to improved efficiency and productivity in the execution of referral activities. Workarounds were reported to provide instant access to patient information thus improving the referral workflow efficiency.

*"WhatsApp is really quick and it's frequently accessible and used by colleagues ." [PT_6]*

For example, in case two where a mandated health information system was not officially adopted for patient referral purposes, HCPs relied on the WhatsApp platform to exchange images, X-rays and other patient-related information. Sharing this information in real-time allowed HCPs to make the necessary diagnosis and treatment decisions at the point of care. As a result, using WhatsApp thus streamlines communication, care coordination and increases the productivity of HCPs. Ultimately, leading to more efficient and effective patient care and services.

Secondly, WhatsApp was used for consultation and seeking second opinion from senior consultants and specialists.

*"WhatsApp is actually becoming quite popular, especially with the onset of Covid-19. When patients have respiratory cases, all respiratory cases, are now potentially Covid cases. The first doctor that sees the patient has to take the history, make an assessment and then they can just post it on WhatsApp on a group that has the internal medicine specialist on it to review the case." [PT_28]*

This flattened the hierarchy in the patient referral pathway by allowing HCPs to directly communicate and consult specialists in real-time, instead of following the referral pathway guidelines which can be time-consuming. This yielded positive outcomes to the referral process in the public hospitals, as non-critical cases could be addressed at lower levels of care and therefore resolving delays and bottlenecks in the referral process. In healthcare settings, specialists possess the medical knowledge which nurses and medical officers lack, and this causes information asymmetries. In resource constraint settings, WhatsApp was beneficial in facilitating and coordinating the transmission of information to handle this constraint. HCPs could seek instant second opinion from specialists who may not work at the same healthcare facility or hospital. This presents opportunities of sharing information and HCPs receiving expert advice which can improve the quality of care rendered to patients.

Thirdly, WhatsApp was also used as a platform to disseminate information and to share knowledge among the community of HCPs and specialists. In case study one, useful information such as journal articles, and medical books and clinical guidelines were shared on WhatsApp and Telegram. Sharing the best practices and educational resources is beneficial because healthcare providers and doctors are improving their knowledge and skills.

*"You know at the moment we're all onto Telegram because Telegram you can share books … It's like WhatsApp but you can share books." [PT_2]*

Positive outcomes for health organisations include continuous professional development for staff, ensuring HCPs stay updated with medical advancements and evidence-based practices.

### *5.2. Improvisations, Design-reality gaps and negative impacts*

The empirical findings demonstrated a relationship between design-reality gaps and workaround practices enacted by HCPs in both case studies. The findings are presented using matrix coding consisting of code references which serve to illustrate instances of data coded, and the relationships between improvisation types and design-reality gaps. In case study one, workaround practices were enacted by HCPs due to gaps related to poor management systems and structures (30 code references), lack of HCPs' involvement (20 code references) in the roll-out of Vula and also due inadequacies of Vula (18 code references) to support referral tasks as shown in appendix A. Additionally, HCPs were enacting workarounds due to the absence of IT policies (25 code references) for regulating the adoption and use of third-party applications in the public hospitals under study. In case study two, there are more gaps observed from the management systems and structures (47 code references), technology (37 code references), information and processes (31 code references each). Other gaps were also observed along the staffing and skills dimension. There are a few gaps observed on the staffing and skills, other resources: objectives and values of health information systems. Improvisations in a form of WhatsApp is observed from the data. Use of WhatsApp for service delivery highlight inefficiencies in the current processes and inadequacies with existing HIS. These workarounds provide critical insights into existing mandated systems and referral workflows, offering a basis for evaluation and improvement. For example, using WhatsApp to share non-sensitive patient information has minimal consequences, whereas sharing sensitive information poses greater risks. Identifying these workarounds can drive process improvement initiatives, helping organisations understand end-user needs and design more effective health information systems. Anticipating workarounds allows for the development of systems that address the real-life needs of healthcare professionals, ultimately enhancing service delivery and achieving referral outcomes. The next section provides a further elaboration of design-reality gaps (DRG) and their effect on referral outcomes as they relate to organisational, patients and healthcare practitioners (as shown in Appendix A).

#### *5.2.1. Poor management structures, implementation of e-health strategies and impacts*

Empirical evidence from both case studies reveals poor management structures due to misalignment between IT and health organisations. Furthermore, there are similarities in how e-health interventions were deployed in a top-down approach in both case studies. There were poor plans of rolling out and implementing e-health strategies to strengthen health service delivery. The top-down approach led to poor integration of HIS in the referral systems and lack of HCPs' lack of awareness of general e-health strategies and policies related to referral processes. In case study one, the HIS was poorly integrated with existing systems, and in case study two, the e-referral application faced integration issues, especially in Hospital A, where HCPs covertly resisted its use. This poor integration resulted in a loss of synergy among HCPs. Additionally, poor management led to inadequate plans for rolling out e-health interventions was reported. Despite the deployed implementation strategies in case study one, the rollout of the e-referral application was poorly conducted, with no guidelines published for post-training and a delayed mandate from the Department of Health, causing HCPs to perceive Vula as non-mandatory. Consequently, the lack of specific guidelines and training manuals weakened the use of the application among HCPs. At the time of this study, the e-referral application was not mandated by the governing department of health in both settings, creating a perception among HCPs that it was optional and not compulsory for facilitating referrals. However, the application was later officially mandated as the formal e-referral tool in the Western Cape Department of Health and it is expected to significantly improve and streamline the referral process [21].

*5.2.2. Lack of end-user involvement and impacts*

There was limited engagement between the IT personnel and HCPs. HCPs from both case studies reported that they were minimally involved in the decision-making process of choosing e-referral applications. While the IT personnel were entrusted with the implementation of Vula, they consulted only senior management and not HCPs as end-users to gather the business and user requirements. There was minimal involvement of the HCPs in the selection process of the application. In addition, in case study one, there was user training conducted at the initial implementation of Vula, where HCPs were introduced to the features of the initial version of the software. However, there was no further involvement of HPCs and follow-up training especially for establishing additional user needs beyond the implementation phase. When such training was conducted, the end-users' needs were sometimes not considered at all and were not implemented in a timely manner. At the time of this study , the lack of user needs resulted in a poor user interface design that constrained HCPs to complete their work and this reduced the effectiveness of the application. As a result, HCPs were adapting and fitting the interface to complete their work. For example, they used text fields to capture additional diagnosis notes for patients, although the text fields were not designed for that purpose. Moreover, some referral messages on Vula were processed with incomplete information, which in turn affected the referral outcomes. This occasionally resulted in an ineffective referral process at the organisational level (hospitals). In some instances, HCPs resorted to augmenting the health information systems with paper-based referrals such as completing referral letters to completement Vula.

*5.2.3. IT Infrastructure and Technological Issues*

HCPs from both cases reported that the IT department was slow in fulfilling their needs. For example, HCPs indicated that IT support did not fully implement the requirements they outlined in the meetings held with management. For example, there were recommendations to fix inadequacies such as the referral message alert tone and to change radio buttons to fields that allow HCPs to capture multiple options for patient information. Furthermore, HCPs from both cases reported that there was lack of IT support with regards to providing guidelines on the use of personal devices for professional work. Other findings show that existing e-health applications worked in silos with other systems that were part of the health organisations [20, 21]. This finding points to lack of integration between Vula and other official systems. The lack of support and lack of integration was a motivator to augmenting or complementing these official systems with other paper-based referral methods or altogether resorting to third-party applications such as WhatsApp. The effect on health organisations, in particular Hospital A in case study one, was that there was limited adoption of the Vula application. In both cases, it was reported that internet service provision from IT was limited and, in some cases, non-existent.

The cost of providing internet-dependent applications was not considered during the implementation phase. HCPs in both cases were therefore left to use personal mobile devices and personal data bundles to complete professional tasks. Due to the cost burden, there was lack of buy-in from some HCPs, and they resorted to improvisation in a form of WhatsApp use. The effects of these are the loss of control for data as the data was stored on personal devices. Using personal devices to store and transmit healthcare data poses several security risks.

IT loses control over data and applications, leading to security vulnerabilities, vendor lock-in, and potential violations of data privacy laws. The IT department therefore did not have control over applications deployed by HCPs. Moreover, there was non-compliance with POPIA (Protection of Personal Information Act) with regards to storing patient data.

*5.2.4. Privacy, security, and ethical issues*

The enactment of various improvisations such as adaption of the existing software applications, use of paper-based methods and WhatsApp, raises privacy and ethical concerns. Although some workaround practices brought about productivity gain from HCPs and health organisations, it posed serious patient safety concerns. In particular, the use of WhatsApp brought about security risks related to unintentional disclosures of patient data. In both case studies, HCPs did not seek consent to take patient photos. As a result, the privacy of patients' information was compromised. In addition, IT policies regulating third-party applications were absent from public hospitals under study. Due to the lack of Shadow IT policies in both cases, health organisations could not manage the adoption and use of third-party applications to ensure patient data privacy, regulatory compliance, and efficient operations of the referral process. In addition, HCPs from both cases reported lack of awareness of referral policies and general e-health strategies. Given this lack of awareness, there was a poor understanding of data security risks. HCPs were therefore prone to ignore relevant policies and procedures, which lead to unintentional disclosures of patient data. Furthermore, lack of policy awareness from among HCPs posed significant security to health organisations. When HCPs are unaware of these policies, they unintentionally enact workarounds practices that compromise patient safety, jeopardize data security and fail to meet compliance standards. From the above findings, it can therefore be argued that workarounds to mandated health applications poses significant security risks. An extended concept on privacy, security and ethical issues was adopted from the adapted design-reality gap model by Makoza and Albertus [26]. Findings show that workarounds to mandated systems raise several ethical concerns. Firstly, the use of WhatsApp by HCPs involves transmission and storage of patient data on a third-party application which is not controlled by the IT department. This increases the risk of data breaches and unauthorized access to sensitive patient information as it is stored on third-party servers and personal devices. Additionally, information silos created by unofficial apps hinder continuity of care and referral outcomes, placing ethical burdens on healthcare providers. This suggests a need for stricter data transmission policies and regulations around Shadow IT within healthcare settings.

## 6. Conclusion

While this research advanced an understanding of how healthcare providers (HCPs) improvise HIS using WhatsApp for facilitating the patient referral process in public hospitals, the research uncovered that the official HIS did not fully streamline the referral workflow as per design objectives. Healthcare providers were resorting to work around practices to complete their referral activities. This finding is supported by several scholars arguing that workarounds result from misfits between systems and work practices of actors [33, 34]. As a result, users institute workarounds to appropriate solutions to these misfits or gaps [2, 35, 36]. This study contributes to the body of knowledge in by providing an explanation that improvisations are result of the gap between the design of the IT artefact and user needs. The study findings from this research are therefore in agreement with these scholars, that there is a need to align Vula's design with HCPs' social context and their specific needs. Developers must address the "design-reality gap" to ensure its effective use. System developers need to pay attention to design-reality gap issues to devise measures to address these gaps. Practitioners and management in healthcare organisations need to recognize that workarounds are common and will occur even though they are non-compliant with existing policies. Given the prevalent use of workarounds in healthcare settings, healthcare organisations should invest in developing effective strategies for managing security risks, prioritizing patient safety, and safeguarding sensitive patient information. Workarounds are an opportunity for innovation and process improvements in health organisations. The results of this research can help public health organisations in dealing with this phenomenon by developing more effective measures, policies, and strategies to address these practices.

**Acknowledgements**
The financial support from the University of Cape Town and Namibian National Commission on Research, Science and Technology (NCRST) is acknowledged.

## Appendix A. Cross-case comparative findings: Improvisations, Design-reality Gaps, and Referral Impacts

| Design-reality Gaps | Description | Improvisation type | Case study 1 | Case study 2 | Negative impacts | | |
|---|---|---|---|---|---|---|---|
| | | | | | **Patient safety** | **HCPs** | **Organisational** |
| **Poor management systems and structures.** | Poor plans for rolling out and implementing e-health strategies to strengthen health service delivery. | Workarounds (Shadow IT) | 30 | 47 | None | Unawareness of e-health strategies and policies. | Poor integration and loss of synergies. |
| **Lack of end-user involvement.** | Lack of user involvement during the selection process of the e-health application. | Fitting<br>Augmenting<br>Workarounds (Shadow IT) | 20 | 1 | None | Inaccurate user requirements lead to inadequate user interface design.<br>Misunderstanding of end-user needs. | Synergy loss and creation of inefficiency. |
| **e-referrals Inadequacies and shortcomings.** | Inadequacies of e-referrals application<br>User interface restricting work of HCPs. | Fitting<br>Augmenting<br>Workarounds (Shadow IT) | 18 | 0 | Compromised patient privacy.<br>Ineffective care rendered to patients. | Incomplete data entry | Data inconsistency and loss. |
| | Software incompatibility with user devices. | (Workarounds (Shadow IT)) | 5 | 12 | | Resistance to e-referrals. | Lack of user buy-in. |
| **IT Infrastructure and technological Issues.** | Limited training on the HIS and Vula Application. | Fitting<br>Augmenting<br>Workarounds (Shadow IT). | 61 | 2 | Loss of control (Data)<br>Unintended use of HIS leading to incorrect information on patients. | None | Uncontrolled vendor dependencies.<br>Unknowledgeable workforce.<br>Lack of user buy-in. |
| | Use of personal devices (BYOD) to share sensitive patient data. | | 19 | 0 | Compromised privacy of patient information. | Lack of control over HCPs devices containing sensitive patient data. | Legislative (POPIA) non-compliance with POPIA. |
| | Slow responsiveness of IT on implementing requests. | Workarounds (Shadow IT), | 5 | 10 | None | Dissatisfaction and limited use of HIS. | Lack of user buy-in. |
| | Lack of integration with official systems. | Augmenting, | 2 | 15 | None | None | Limited adoption of the Vula application in some hospitals. |
| | Lack of internet access. | Augmenting<br>Workarounds (Shadow IT), | 17 | 20 | None | Limits HCPs to connect and sharing information with other HCPs.<br>Inability to connect to the HIS. | None |
| **Absence of IT policy and Lack of policy awareness** | HCPs lack awareness of referral policies.<br>Lack of IT policies to manage risks of Workarounds (Shadow IT). | Fitting<br>Augmenting<br>Workarounds (Shadow IT). | 25 | 33 | Compromised patient privacy.<br>Ineffective care rendered to patients | Non -compliance to referral protocols /policies.<br>Poor understanding of data security breaches<br>Incomplete data entry. | Data inconsistency and loss.<br>Lack of control for the IT department. |